\documentclass{article}
\usepackage{amssymb}
\usepackage{graphicx}
\begin{document}

%\begin{frontmatter}
\title{On the Security of Y-00 under Fast Correlation and Other Attacks on the Key}
\author{Horace P. Yuen \footnote{yuen@ece.northwestern.edu} and
Ranjith Nair
 \\Center for Photonic Communication and Computing\\
Department of Electrical and Computer Engineering\\
Department of Physics and Astronomy\\
Northwestern University, Evanston, IL 60208} \maketitle

\begin{abstract}
The potential weakness of the Y-00 direct encryption protocol when
the encryption box ENC is not chosen properly is demonstrated in a
fast correlation attack by S.~Donnet et al in Phys. Lett. A 356
(2006) 406-410.  In this paper, we show how this weakness can be
eliminated with a proper design of ENC. In particular, we present
a Y-00 configuration that is more secure than AES under
known-plaintext attack. It is also shown that under any
ciphertext-only attack, full information-theoretic security on the
Y-00 seed key is obtained for any ENC when proper deliberate
signal randomization is employed.
\end{abstract}
%\begin{keyword}
%Quantum cryptography
%\PACS 03.67.Dd
%\end{keyword}
%\end{frontmatter}

\section{Introduction}

The quantum-noise based direct encryption protocol Y-00, called
$\alpha\eta$ in our earlier papers [1-6], was repeatedly
misrepresented in previous criticisms, but that situation has
apparently changed with our recent papers [7-9]. For the first
time, a meaningful attack on Y-00 type protocols beyond exhaustive
search has been developed in \cite{donnet}. A fast correlation
attack (FCA) was presented that was shown to succeed by
simulations for moderate signal levels when the ENC box in Y-00 is
a LFSR (linear feedback shift register) of a few taps and length
up to 32. Even though such Y-00 is already insecure against what
we call assisted brute-force search \cite{nair06} due to the small
seed key size $|K| \leq 32$, such FCA is of interest as it brings
forth the whole issue of Y-00 seed key security against similar
and other attacks.

The attack in \cite{donnet} is geared toward only the experiment
reported in \cite{optlett03}. We have emphasized all along
\cite{yuen04,ptl05,spie05} that the use of LFSR in the reported
experiments was just for proof of principle demonstration, that
the ENC box must be chosen appropriately in a final design, and
that other techniques need to be deployed for proper security. To
quote from \cite{spie05}, ``Similar to encryption based on
nonlinearly combining the LFSR's, Eve can launch a correlation
attack using the following strategy: $\ldots$ many of the LFSR's
could be trivially attacked.'' Thus, we were aware of the possible
weakness of some ENC and in particular of FCA type attacks.
Indeed, Hirota and Kurosawa \cite{hirota06} have already desribed
a counter-measure to FCA via a ``keyed mapper'', the incorporation
of which in ASK-signal Y-00 \cite{hirota05} has been developed and
is being tested. Generally speaking, it is important to study
LFSR-based Y-00 despite its possible weakness, because LFSR is a
practically convenient choice in various applications similar to
the situation in standard cryptography.

In this paper, we first briefly describe general attacks on the
Y-00 seed key as a problem of decoding in real noise -- a
viewpoint which includes all FCA's. For both ciphertext-only
attacks (CTA) and known-plaintext attacks (KPA), we show that Y-00
may be considered as a classical stream cipher, the ENC box, with
real physical noise added on top. We comment on the possible
defenses involving just a properly chosen LFSR, or an added keyed
mapper, or with a keyed connection polynomial for the LFSR. We
describe an AES-based Y-00 that is more secure against KPA than
AES (Advanced Encryption Standard) alone, in the sense that if it
is broken then AES is also broken but not the other way around.
The practical security advantage of such AES-based Y-00 will be
indicated. Finally, for CTA, we show that Deliberate Signal
Randomization (DSR) introduced in \cite{yuen04} provides full
information-theoretic security on the Y-00 seed key for any ENC.
We hope that these results would establish beyond doubt that Y-00
is an important cryptosystem to consider in theory and in
practice.

\section{Attacks on Y-00 seed key}
Consider the original quantum-noise randomized cipher Y-00
\cite{prl,yuen04} as depicted in Fig.~1. Alice encodes each data
bit into a $2M-$ary phase-shifted coherent state in a qumode of
energy $\alpha^2_0$. A seed key $K$ of bit length $|K|$ is used to
drive a conventional stream cipher ENC to produce a running key
$K'$ that is used to determine, for each qumode carrying the bit,
which pair of antipodal coherent states, referred to as a basis,
is to be used as a binary phase-shift keying (BPSK) signal set for
Bob. With a synchronous ENC at the receiver, Bob discriminates the
BPSK signals for each qumode by an appropriate receiver. With a
differential (DPSK) implementation
\cite{prl,yuen04,optlett03,ptl05,pra05,spie05}, there is no need
to phase lock between Alice and Bob as is true in ordinary
communications.
\begin{figure*} [htbp]
\begin{center}
\rotatebox{0} {
\includegraphics[scale=0.5]{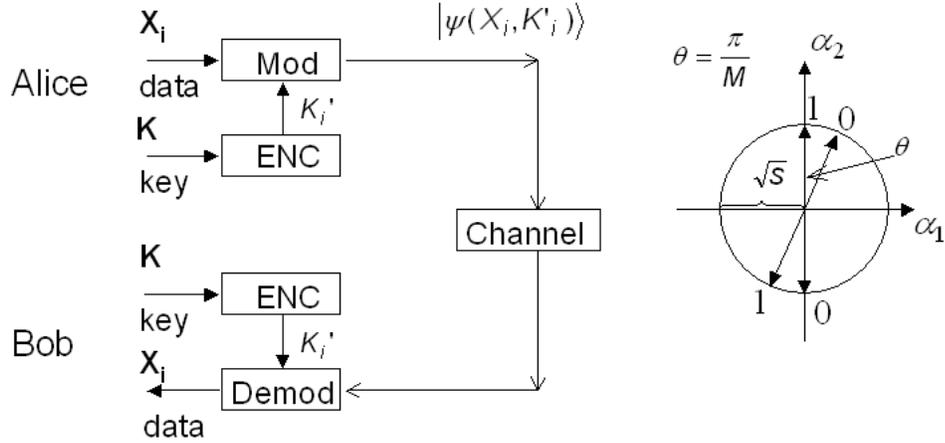}}
\caption{Left -- Overall schematic of the Y-00 encryption system.
 Right -- Depiction of two of $M$ bases with interleaved logical
bit mappings.}
\end{center}
\end{figure*}
\newline\newline
The optimum quantum receiver performance for both Bob and Eve is
the same as in the non-differential case in principle, the
differential implementation being a practical convenience. Even
with a full copy of the quantum state granted to Eve in our KCQ
approach of performance analysis
\cite{yuen04,pla05,yuen05qph,nair06}, security on the data is
nearly perfect when the seed key induced correlation is neglected
\cite{prl}. Generally, it is a horrendous problem with yet no
solution for meaningfully quantifying the data security of a
symmetric-key cipher. In current practice, it is assumed that CTA
on the data is not a problem if $|K|$ is ``large'', and attention
is focused on KPA on the key.

For conventional or standard \cite{yuen05qph,nair06} ciphers, the
key is usually completely protected from CTA for uniformly random
data. This is, however, not the case for the bare Y-00
\cite{yuen04,pla05,yuen05qph,nair06}. In this paper, we address
both CTA and KPA on the Y-00 seed key, the (classical) ciphertext
being obtained from some quantum measurement on the qumodes
assumed to be in Eve's possession. It is seen from Fig.~1 that a
CTA or KPA on the Y-00 seed key is equivalent to the corresponding
attack on the standard stream cipher ENC with its output stream
observed in noise resulting from the coherent state randomization
of the signal phase. Thus, it is equivalent to a CTA or KPA on the
ENC alone as a stream cipher but with noise on top. The connection
between the running key bits $K'$ and the basis, called the
``\emph{mapper}'' \cite{pra05,spie05,hirota06}, a crucial
component of Y-00, and the noise effect on $K'$ are described in
\cite{spie05,donnet}. In a FCA on a conventional stream cipher
composed of, say, a nonlinear combination of the outputs of a bank
of $m$ LFSR's, one focuses on one LFSR $L_i$ at a time and looks
for correlation between the final stream cipher output $K'$ and
the output $k'_i$ of $L_i$. Thus, even though the complete cipher
is nonrandom, $K'$ constitutes a noisy observation of $k'_i$ from
which a good estimate of $k'_i$ may perhaps be obtained. Such a
divide-and-conquer strategy can be repeated to yield all the keys
$k_i$ for each $L_i$. For Y-00, there is real noise from the
coherent states, but a similar FCA can be launched if there is a
significant correlation between $K'$ and the observed $2M$-ary
signal, as obtained, say, by heterodyning.

In general, attack on the Y-00 seed key is \emph{exactly} a
decoding problem on a memoryless channel for both CTA and KPA.
This can be seen by regarding the seed key as information bits and
the observed sequence of $2M$-ary signals translated by the mapper
to $K'$ as the codeword, with independent coherent-state noise for
each qumode so that the memoryless channel alphabet has size
$\log_2 2M$ in a CTA and $\log_2 M$ in a KPA. Note that this code
from ENC, as in the case of AES, could be nonlinear with no useful
linear approximation, making linear decoding not a viable attack.
It is not known whether information-theoretic security may be
obtained in Y-00 for a properly designed ENC, i.e., whether a
(decoding) algorithm may be found that would succeed in
determining the seed key with some nonvanishing probability
\cite{yuen05qph,nair06}. And there is the further question, if
such an algorithm exists, of its complexity as the general
syndrome decoding of even a linear code is exponential. In
contrast, for KPA on standard ``nondegenerate'' nonrandom ciphers,
the key is actually uniquely determined at a bit length $n_1=n_d$,
the nondegeneracy distance \cite{yuen05qph,nair06} which is often
not very long. Thus, such cipher has no information-theoretic
security against KPA, although there is still the problem of
attack complexity in finding $K$ that may allow complexity-based
security which can be practically as good as information-theoretic
security \cite{yuen05qph}. The key point in this connection is
that randomization introduces real noise that is otherwise absent
in a nonrandom cipher, signifying its role in adding security to
KPA.

For standard stream ciphers built upon LFSR's, the class of FCA
described above is powerful enough to break them  for sufficiently
long observed length $N$ of the output. However, the complexity of
all known FCA algorithms is exponential in either the memory
needed or the number  $t$ of tap coefficients in the LFSR [13].
Thus, practically there are LFSR-based stream ciphers that are not
broken by any known attack when the LFSR length $|K|$ and $t$ are
sufficiently large. Shorter LFSR's or ones with long $|K|$ and
with few taps are more convenient and cheaper to use in practice,
but are vulnerable to computationally intensive but feasible
attacks. If such LFSR is used in the ENC in Y-00, the cipher
becomes vulnerable even for moderate signal level if long enough
$N$ is employed when that does not lead to an undue increase in
memory required. For the $|K|=32$ single LFSR case reported in
\cite{donnet}, only $N=1500$ is needed in a CTA to undermine the
system at the signal level $\alpha^2_0 \sim 1.5\times 10^4$,
roughly the numbers used in \cite{optlett03}. The
convolutional-code based algorithm chosen in \cite{donnet} is not
suited to attacking long $|K|$ LFSR with a few taps, and thus
would not be able to break the $|K|=4400$ and $t=3$ LFSR used in
our system in \cite{pra05,corndorf04}. However, a different FCA
would no doubt be able to break that system, such as those
designed for small $t$.

\section{Defenses Against Fast Correlation Attacks}

We have already observed that one may use practical LFSR that
resists known FCA in the ENC of Y-00. There are many other ways to
defeat such and even more general attacks on the Y-00 seed key, as
we will discuss in the rest of this paper.

First, a properly designed deterministic mapper that determines
the $2M-$ary signal from the running key $K'$ would spread the
noise into the different bit positions of $K'(m)$, increasing the
minimum complexity of attacks. The mapper may also be keyed, e.g.,
the mapper function may be chosen for each qumode from the running
key $K_m'$ from another ${ENC}_m$ with another seed key $K_m$.
This results in a product cipher of ENC in noise and $ENC_m$, for
which no obvious modification of the FCA can be made that does not
involve exponential search over $K_m$. In particular, one cannot
plot Fig.~3 in \cite{donnet} which is the basic starting point of
their attack. This defense has already been proposed
\cite{hirota06}, although there is ``correlation immunity'' for
such ciphers only under an approximation.

Secondly, the connection polynomial in the LFSR can be keyed, i.e.,
chosen randomly from an exponential number of possibilities. The
known FCA's on LFSR all require knowledge of the LFSR connection
polynomial. In a future paper, we will present information-theoretic
analysis of the effect of a keyed connection polynomial. Such
ciphers can clearly be implemented in software, and to a
considerable extent in hardware with field programmable logic, thus
retaining much of the convenience of LFSR in practical applications.
We do not believe information-theoretic security can be obtained
this way, but it may greatly increase the complexity of at least FCA
type attacks, thus providing useful practical security in some
situations.

Thirdly, we now give an ENC design for Y-00 that leads to
exponential complexity for CTA according to current knowledge, and
more security that AES for KPA generally. Consider the ENC of
Fig.~2 where a bank of $m$ parallel AES in a stream cipher mode is
used to provide the $m=\log_2 M$ bits running key segment $K'(m)$
which determines, through the mapper, the basis of a qumode.
Typically in our previous experimental demonstrations, $m \sim 10$
and $|K|$ is in the thousands. Thus each $K_i$ may be readily
chosen to be of 256 bits. Under heterodyne or any other quantum
measurement by Eve, the result is a noisy version of $K'(m)$ with
independent coherent-state induced randomization for each qumode.
According to the present state of knowledge, no KPA on AES is
better than exhaustive (exponential) search \cite{stinson}. Even
in a divide-and-conquer type attack as in FCA, so that a single
AES is to be considered, one needs to deal with the KPA problem of
artifical noise from such strategy with the addition of
\emph{real} coherent-state noise, in a CTA on the Y-00 seed key.
Let $N_1$ be the length of the qumode sequence used for the
attack, so that Eve may parallelize $N/{N_1}$ attacks
simultaneously from the total length $N$. It is clear that even
without noise, the attack complexity remains exponential for any
realistic $N \leq 2^{80}$ and any $N_1$. In a KPA, the comparison
is to be made with the same $N_1$ for no parallelization. Thus,
Y-00 is equivalent to AES in a stream cipher mode with output
observed in noise, thus harder than AES alone which does not have
the decoding in noise problem. In particular, it is easily seen
that if the Y-00 in the configuration of Fig.~2 can be broken,
then each $AES_i$ itself can be broken.
\begin{figure*} [htbp]
\begin{center}
 {
\includegraphics[width=4.5in,height=2in]{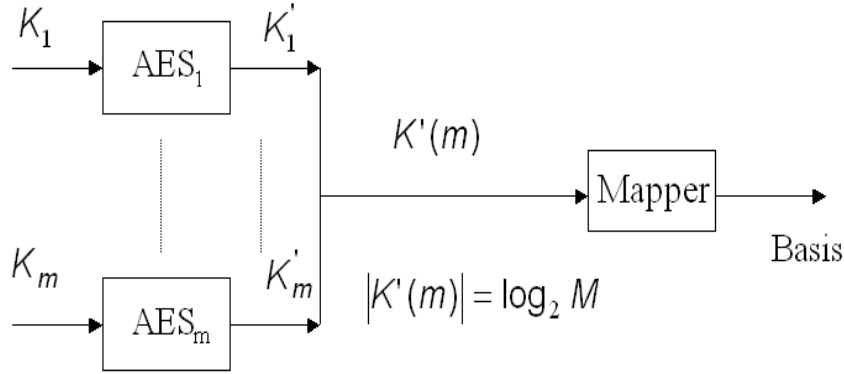}}
\caption{Parallel AES for ENC in Y-00}
\end{center}
\end{figure*}
\newline
The question arises as to what constitutes a fair comparison
between a given stream cipher ENC versus Y-00 on top of ENC. A
different configuration was given for ENC in \cite{nair06}, where
a single classical stream cipher (say AES) is used without
parallelization but is adjusted to give the same clock rate for
encrypting each data bit. The present scheme appears simpler in
principle and more secure in practice when AES is used in ENC,
because the functionality of multiple AES in parallel cannot be
replaced by a single AES. However, with such parallelization for
maintaining the same clock rate as AES (or ENC alone), the
question arises as to whether the added security from Y-00 can be
obtained from, say, nonlinearly combining the parallel AES's. This
question cannot be answered until security is precisely defined
and quantified. However, it may be observed in this connection
that there is no known attack developed for AES observed \emph{in
noise}, and the intrinsic nonlinearity of AES renders all known
decoding attacks inapplicable.

The major qualitative advantage of Y-00 \cite{yuen05qph,nair06}
compared to a standard nonrandom cipher is that the quantum noise
automatically provides high speed true randomization not available
otherwise, thus giving it a different kind of protection from
nonrandom ciphers. Furthermore, one has to attack such
physics-based cryptosystem at the communication line with physical
(measurement) equipment, which is not available to everyone at
every place, whereas one only needs to sit at a computer terminal
to attack conventional ciphers. In this connection, it may be
mentioned that the high rate heterodyne attack needed on Y-00 is
currently not quite technologically feasible, though it may be in
the not-too-far future.

Y-00 can be employed to realize these benefits not available
otherwise. However, if it is intrinsically less secure than
conventional ciphers, its utility would be in serious doubt. The
configuration of Fig.~2  shows this is not the case -- it can in
fact be more secure that ENC or $AES_i$ by itself. There is also
no known attack applicable to AES in noise.

\section{Deliberate Signal Randomization}

In contrast to a nondegenerate nonrandom classical cipher for
which the key is completely protected in the information-theoretic
sense against CTA when the data is uniformly random
\cite{yuen05qph,nair06}, there is little distinction between CTA
and KPA for the bare Y-00. Only a factor of 2 in the per qumode
alphabet size is obtained in KPA versus CTA as indicated above,
and expounded in \cite{nair06}. The question arises as to whether
full information-theoretic security against CTA can be restored by
modifying the bare Y-00.  The authors of \cite{donnet} appear to
be pessimistic on the possibility of achieving this. To quote:
``While randomization methods might increase the security level,
it remains to be seen if they will provide perfect secrecy.'' In
the following, we show how this is possible with Deliberate Signal
Randomization (DSR) independently of the mechanism of running key
generation.

The reason why the seed key cannot be attacked in CTA is clear for
an additive stream cipher with uniformly random data. The
``channel'' between the seed key and the output observation has
zero capacity due to the data which acts as random noise. In
particular, it is clear that no FCA can be launched. The
coherent-state noise in Y-00 is not big enough for high signal
level to produce a similar effect. However, further randomization
may in principle be produced to achieve this end, both classically
and quantum mechanically.

Since the coherent-state noise in Y-00 can in principle be
replaced, in an equivalent classical system, by deliberate
randomization of the classical signal from Alice as we have
repeatedly emphasized \cite{yuen04,pla05,yuen05qph,nair06}, we
first consider this classical situation. Let $\theta_s$ be the
signal point on the circle of Fig.~1, $x$ the data bit, $k'$ the
running key segment that determines the basis. Before deliberate
or noise randomization, $\theta_s(x,k')$ is uniquely determined by
$x$ and $k'$. From $\theta_s$ one randomizes it to $\theta_r$
according to a probability density $p(\theta_r|\theta_s)$. We use
continuous $\theta$'s here but the argument is identical for
discrete $\theta$'s. More generally, let $\theta$ be Eve's
observed signal point, so that $\theta=\theta_r$ in a classical
noiseless system with deliberate randomization. Then,
\begin{equation}\label{pdf}
p(\theta|x,k')=\int
p(\theta|\theta_r)p(\theta_r|\theta_s(x,k'))\mathrm{d} \theta_r.
\end{equation}

In the classical noiseless case with just signal randomization,
$p(\theta|\theta_r)=\delta(\theta-\theta_r)$, the BPSK signal may
be correctly discriminated when the observed $\theta$ falls within
the half-circle centred around $\theta_s$. Thus we pick
$p(\theta_r|\theta_s)$ to be the uniform distribution on the
half-circle with midpoint $\theta_s$. If $x$ is uniformly random,
then from (1)
\begin{equation} \label{pdf2}
p(\theta|k')=\frac{1}{2} \sum_{x=0,1} p(\theta|x,k')
\end{equation}
is the uniform distribution on the full circle independent of
$k'$. This proves the observation of $\theta$ to Eve yields no
information at all on $k'$. In other words, Eve's channel on $k'$
has zero capacity from DSR and uniformly random data which acts as
added noise unknown to her, similar to a nondegenerate nonrandom
stream cipher.

For coherent-state noise described in the wedge approximation
\cite{pla05,nair06}, whereupon a heterodyne or phase measurement
the observed $\theta$ is taken to be uniformly distributed within
a standard deviation around $\theta_r$ and zero outside, the same
$k'-$independence for $p(\theta|k')$ obtains when $\theta_r$ is
chosen in a discrete number of positions for given $\theta_s$ so
that $p(\theta_r|\theta_s)$ fills out a uniform half-circle again.
We have assumed an integral number of wedges would do this, which
can be guaranteed by choice of the signal level $\alpha_0$. Going
beyond the wedge approximation, one needs to determine the
function $p(\theta_r|\theta_s)$ in (1) for a coherent state/fixed
measurement $p(\theta|\theta_r)$ so that $p(\theta|x,k')$ is
uniformly distributed in a half-circle, where $p(\theta|\theta_r)$
is obtained from Eve's optimal individual qumode quantum
measurement. In this case, there is the problem that the resulting
error probability for Bob may be higher than the designed level
even with knowledge of the seed key $K$. In principle, this
problem can be handled in one of two different ways without
affecting the data security as measured by the Shannon limit
\cite{yuen05qph,nair06}.

First, one may increase $S$ and correspondingly $M$ while
maintaining the same Y-00 random cipher characteristic $\Gamma =
M/{\pi\sqrt{S}}$ defined in \cite{nair06}. Doing so will make the
tail of the probability distribution that causes Bob's error
arbitrarily small. Indeed, in the classical limit $S \rightarrow
\infty, M \rightarrow \infty, M/{\sqrt{S}} \rightarrow \pi\Gamma$,
a constant, the error vanishes. A second way is to employ an error
correcting code for Bob and randomize the entire codeword of
$n$-bits in a correlated fashion in the signal space
$\mathcal{C}^n$, where $\mathcal{C}$ is the coherent-state circle
in $\mathbb{R}^2$. This is done by moving the $n$-bit codewords
within mutually exclusive but jointly exhaustive regions that fill
the entire signal space $\mathcal{C}^n$, similar to the filling of
the circle $\mathcal{C}$ in the one-bit case. Detailed
quantitative treatment of these will appear elsewhere.

Note that Y-00 is only a random cipher for a given quantum
measurement, it is not a random quantum cipher. See \cite{nair06}.
A convenient way to make it a quantum random cipher is to
randomize the parameter $\theta_s$ to $\theta_r$ that determines
the quantum state $\rho(\theta_r)$ to be transmitted. The
resulting output state is then, analogous to (1),
\begin{equation} \label{state}
\rho(x,k')=\int
\rho(\theta_r)p(\theta_r|\theta_s(x,k'))\mathrm{d}\theta_r.
\end{equation}
It may be seen from (3) that by uniformly randomizing $\theta_s$
as above, for any state modulation $\rho(\theta_r)$, the output
quantum state itself is independent of $k'$ upon averaging over
$x$ as before. Thus, such quantum DSR would protect the key
against CTA with the most general joint (quantum measurement)
attack. In this case, there is generally a larger probability of
error that Bob would decide on $x$ incorrectly as compared to no
DSR, similar to the specific coherent state case under heterodyne
attack. One of the above two approaches in the fixed measurement
case can be similarly employed to bring the error down to any
desired level.

It may be noted that the deployment of full DSR just described
above is practically difficult at present if only because high
speed random numbers are needed. On the other hand, it may be
possible to delve into the qumode sequence to take advantage of
the randomization inherent in such sequence for selected
deliberate randomization while providing essentially the same
overall result. Detailed treatment of concrete DSR on Y-00 will be
given elsewhere.

\section{Conclusion}

We have shown that Y-00 can be designed to be secure against fast
correlation attacks including that of ref. \cite{donnet}, and that
it can be configured to be more secure than AES while retaining
the same high speed and its advantage as a physics-based cipher.
We also prove the full information-theoretic security of Y-00 with
proper deliberate signal randomization against ciphertext-only
attacks. Quantitative security against known-plaintext attacks, as
in the case of conventional ciphers, is a difficult, open, and
important area of research.

\section{Acknowledgements}
We would like to thank E.~Corndorf, G.~Kanter, P.~Kumar, and
C.~Liang for useful discussions. This work has been supported by
DARPA under grant F30602-01-2-0528 and AFOSR grant
FA9550-06-1-0452.

\bibliographystyle{elsart-num}

\end{document}